\documentclass[twocolumn]{aastex631}

\usepackage{CJK}
\usepackage{bm}

\shorttitle{Coevolution of SMBHs and massive hosts}
\shortauthors{Hu et al.}

\graphicspath{{./}{figures/}}

\definecolor{HHcolor}{rgb}{0.93,0.57,0.13}
\definecolor{HHcolor2}{rgb}{0.5,0.1,0.5}

\begin{document}
\begin{CJK*}{UTF8}{gbsn} 
\title{Supercritical growth pathway to overmassive black holes at cosmic dawn:\\
coevolution with massive quasar hosts}

\correspondingauthor{Haojie Hu(胡豪杰)}
\email{hhj\_pku@pku.edu.cn}

\author[0000-0003-3143-3995]{Haojie Hu}
\affiliation{Kavli Institute for Astronomy and Astrophysics, Peking University, 5 Yiheyuan Road,  Haidian District, Beijing, 100871, PRC}
\affiliation{Department of Astronomy, School of Physics, Peking University, 5 Yiheyuan Road,  Haidian District, Beijing, 100871, PRC}

\author[0000-0001-9840-4959]{Kohei Inayoshi}
\affiliation{Kavli Institute for Astronomy and Astrophysics, Peking University, 5 Yiheyuan Road, Haidian District, Beijing, 100871, PRC}

\author[0000-0003-3633-5403]{Zolt\'an Haiman}
\affiliation{Department of Astronomy, Columbia University, New York, NY 10027, USA}

\author[0000-0002-1044-4081]{Wenxiu Li}
\affiliation{Kavli Institute for Astronomy and Astrophysics, Peking University, 5 Yiheyuan Road,  Haidian District, Beijing, 100871, PRC}
\affiliation{Department of Astronomy, School of Physics, Peking University, 5 Yiheyuan Road,  Haidian District, Beijing, 100871, PRC}

\author[0000-0001-9185-5044]{Eliot Quataert}
\affiliation{Department of Astrophysical Sciences, Princeton University, Peyton Hall, Princeton, NJ 08544, USA}

\author[0000-0003-2309-8963]{Rolf Kuiper}
\affiliation{Fakult\"at f\"ur Physik, Universit\"at Duisburg-Essen, Lotharstra$\ss$e 1, 47057 Duisburg, Germany}

\begin{abstract}
Observations of the most luminous quasars at high redshifts ($z > 6$) have revealed that the largest supermassive black holes 
(SMBHs) at those epochs tend to be substantially overmassive relative to their host galaxies compared to the local relations, 
suggesting they experienced rapid early growth phases. We propose an assembly model for the SMBHs that end up in rare 
massive $\sim10^{12}~M_{\odot}$ host halos at $z \sim 6-7$, applying a kinetic feedback prescription for BHs accreting above the Eddington rate, 
provided by radiation hydrodynamic simulations for the long-term evolution of the accretion-flow structure. The large inflow 
rates into these halos during their assembly enable the formation of $>10^9~M_{\odot}$ SMBHs by $z \sim 6$, even starting from stellar-mass 
seeds at $z \sim 30$, and even in the presence of outflows that reduce the BH feeding rate, especially at early times. This mechanism 
also naturally yields a high BH-to-galaxy mass ratio of $> 0.01$ before the SMBH mass reaches $M_{\rm BH} > 10^9~M_{\odot}$ by $z \sim 6$. 
These fast-growing SMBH progenitors are bright enough to be detected by upcoming observations with the James Webb Space 
Telescope over a wide range of redshift ($7 < z < 15$), regardless of how they were seeded.
\end{abstract}

\keywords{Supermassive black holes (1663); Quasars (1319); High-redshift galaxies (734)}

\section{Introduction} \label{sec:intro}

Observations of active galactic nuclei (AGN) have revealed the presence of supermassive black holes (SMBHs) harbored in the centers of 
galaxies at a wide range of redshift, $z\sim 0-7$ \citep{Mortlock2011,Wu2015,Banados2018,Matsuoka2018a}, 
offering stringent constraints on the formation of such massive monsters \citep{Inayoshi2020,Volonteri2021}. 
The empirical correlations between the mass of SMBHs ($M_{\rm BH}$) and the host galaxy properties 
(e.g., bulge mass $M_{\rm bulge}$ and total galaxy stellar mass $M_\star$) in the local universe are expected to be an outcome 
of ``BH-galaxy coevolution" over cosmic time \citep[e.g.,][]{kormendy2013coevolution,Reines2015}, but the origin of this coevolution remains 
an unsolved puzzle in the framework of galaxy formation.

Toward higher redshifts ($z\sim 6$), the mass ratio of $M_{\rm BH}/M_\star$, where $M_\star$ is approximated by the dynamical mass 
$M_{\rm dyn}$ measured from gas kinematics using, e.g., ALMA, appears to be significantly elevated compared to the local value \citep{Wang2010,Wang2013},
suggesting that the most massive SMBHs at $z\ga 6 $ got a head start over the growth of their host galaxies.
However, those quasars represent the tip of the iceberg of the high-$z$ BH population found in shallow surveys (e.g., SDSS)
rather than the underlying populations detected in deeper surveys \citep[e.g., Subaru HSC;][]{Matsuoka2016,Onoue2019,Izumi2021}. 
The mass ratio for the bulk population inferred from current observations seems consistent with the local value within errors owing to the intrinsic scatter and 
the strength of various systematic uncertainties \citep{Li2022}. 
Further improvements of the mass measurements and exploration of less luminous quasars are required to better understand
the physical origin of the BH-galaxy coevolution.

The early coevolution problem has been extensively studied by theoretical work, especially with galaxy formation simulations.
However, as described in \citet{Habouzit2022}, the redshift dependence of $M_{\rm BH}/M_\star$ shows a great diversity 
depending on how stellar and AGN feedback processes are treated as subgrid physics that are unresolved in large-scale simulations.
As a result, most galaxy simulations \citep[e.g.,][]{Zhu2020,Valentini2021}
yield BH populations with $M_{\rm BH}/M_\star<0.01$ at $z\sim 6$ that start to grow in mass when the host galaxies become sufficiently massive.
In contrast, radiation hydrodynamical (RHD) simulations resolving nuclear scales suggest that gas supply from galactic scales
promotes rapid mass accretion onto BHs \citep{jiang2014,sadowski2015,Inayoshi2016,Toyouchi2021} and 
the transient super-Eddington accretion mode naturally yields a mass ratio of $M_{\rm BH}/M_\star>0.01$ higher than the local value 
\citep{Inayoshi2022}. Recently, in \citet{Hu2022}, we performed a series of long-term RHD simulations for super-Eddington accreting flows onto a BH
and proposed a subgrid feedback model associated with outflows, which can be applied to large-scale cosmological simulations.

In this paper, we incorporate this feedback model for super-Eddington accreting BHs into a Monte Carlo merger tree based 
model for the assembly of the first massive BHs observed in high-redshift quasars. 
In this model, almost all nuclear BHs grow faster than their host galaxies at early times even with strong outflows, and 
reach the overmassive region in the BH-galaxy mass diagram.

\begin{figure}
\centering
\includegraphics[scale=0.30]{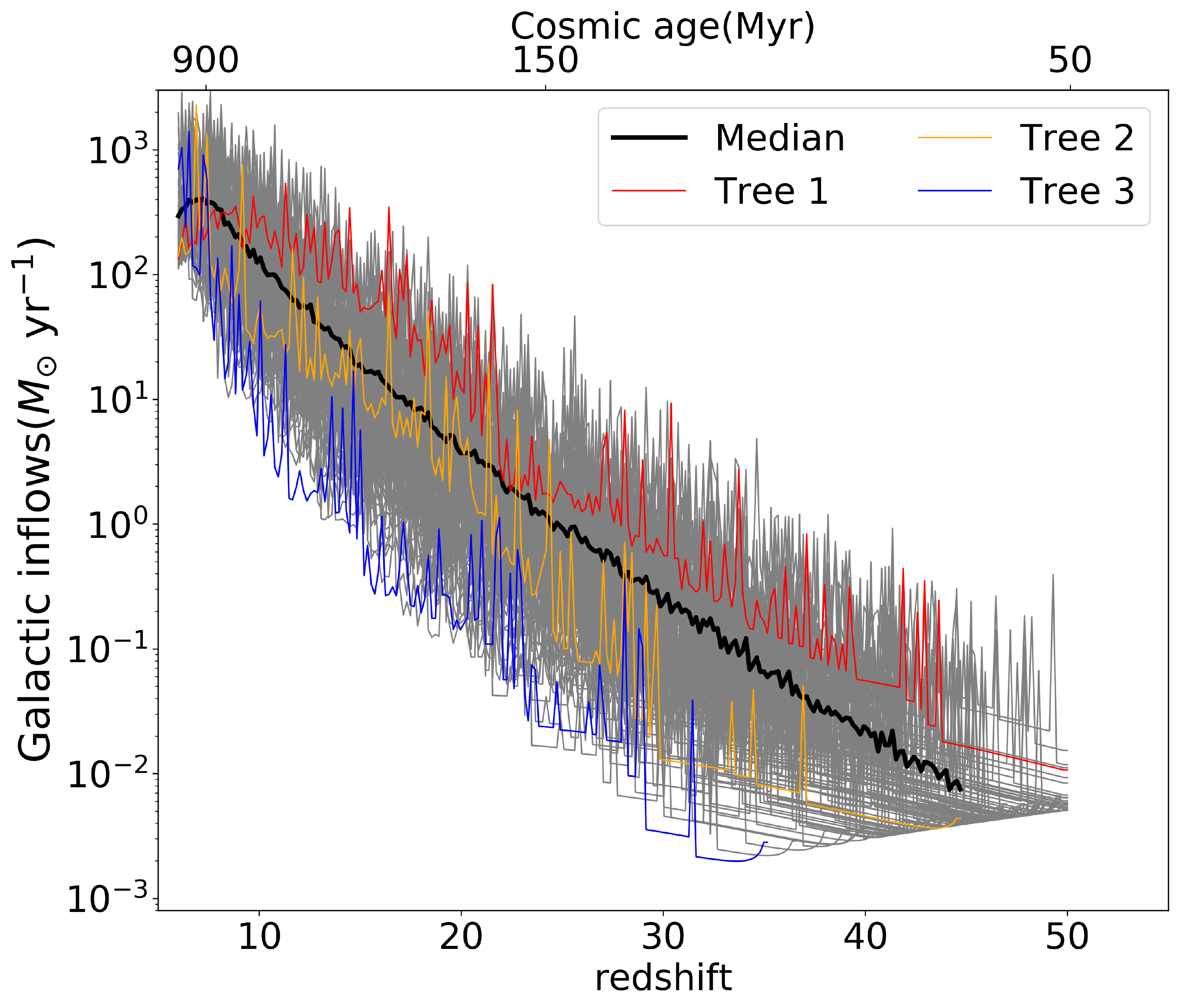}
\caption{Galactic mass inflow rate as a function of redshift based on the assembly history of DM halos 
that end up in high-z quasar host galaxies with $M_{\rm h}=10^{12}~M_\odot$ at $z=6$ \citep{Li2021}. Among the $10^4$ merger 
trees, three representative cases (red, orange, and blue) are highlighted and the median inflow rate is overlaid (black). }
\label{fig:mergertree}
\end{figure}

\section{Methodology} \label{sec:method}

In \citet[][]{Hu2022}, we study the long-term evolution of the global structure of accretion flows onto a BH at rates substantially higher than 
the Eddington value $\dot{M}_{\rm Edd}[\equiv L_{\rm Edd}/(0.1c^2)]$, performing two-dimensional axisymmetric RHD simulations
that cover a computational domain from $r_{\rm min}=3~r_{\rm Sch}$ to $r_{\rm max}=1500~r_{\rm Sch}$, 
where $L_{\rm Edd}$ is the Eddington luminosity and $r_{\rm Sch}$ is the Schwarzschild radius of the BH \citep[see more details in][]{Hu2022}. 
When the gas supply rate from larger radii is substantially higher than the Eddington value, i.e., $\dot{M}_0 \gg \dot{M}_{\rm Edd}$,
the radiative luminosity is reduced owing to photons trapped within the dense flow, but strong bipolar outflows are launched within the dense trapping region.
The numerical results show that the mass inflow rate decreases owing to the outflows toward the center as $\propto r^p$ with an index of $p\sim 0.5-0.7$ and 
thus only a small fraction of the gas supply is swallowed by the central BH.
Motivated by the simulation results, we adopt a BH mass growth model as
\begin{equation}
\dot{M}_{\rm BH} = \dot{M}_0 \left(\frac{r_{\rm min}}{r_{\rm tr}}\right)^p~~~~{\rm if}~r_{\rm min}\leq r_{\rm tr}
\label{eq:inflow}
\end{equation}
and $\dot{M}_{\rm BH}=\dot{M}_0$ otherwise, where $r_{\rm min}$ is set to the radius of the inner-most stable circular orbit for a non-rotating BH
and $r_{\rm tr}~[\equiv 5 \dot{M}_0 r_{\rm min}/(3 \dot{M}_{\rm Edd})]$ is the photon trapping radius. 
The reduction of the inflow rate is generally found in most previous simulations of radiatively inefficient accretion flows
($p\sim 0.5-1$; see \citealt{Stone1999,Igumenshchev2003,Yuan2014}).
The power-law index $p\sim 0.5-0.7$ seen in \citet{Hu2022} is relatively smaller than that predicted in the convection-dominated 
accretion flow \citep[$p\simeq 1$;][]{Narayan2000,Quataert2000,Abramowicz2002} but is consistent with
(magneto-)hydrodynamical simulations that cover a wide range of spatial scales \citep[$0.5\la p \la0.7$; ][]{Pen2003,Yuan2012,Ressler2020, Guo2020}.
Since the value of $p$ characterizing the outflow strength depends on various simulation setups,
we adopt $p=0.5$ as our fiducial case but also study the dependence of the choice on the resultant BH growth.

\begin{figure*}
\centering
\includegraphics[scale=0.34]{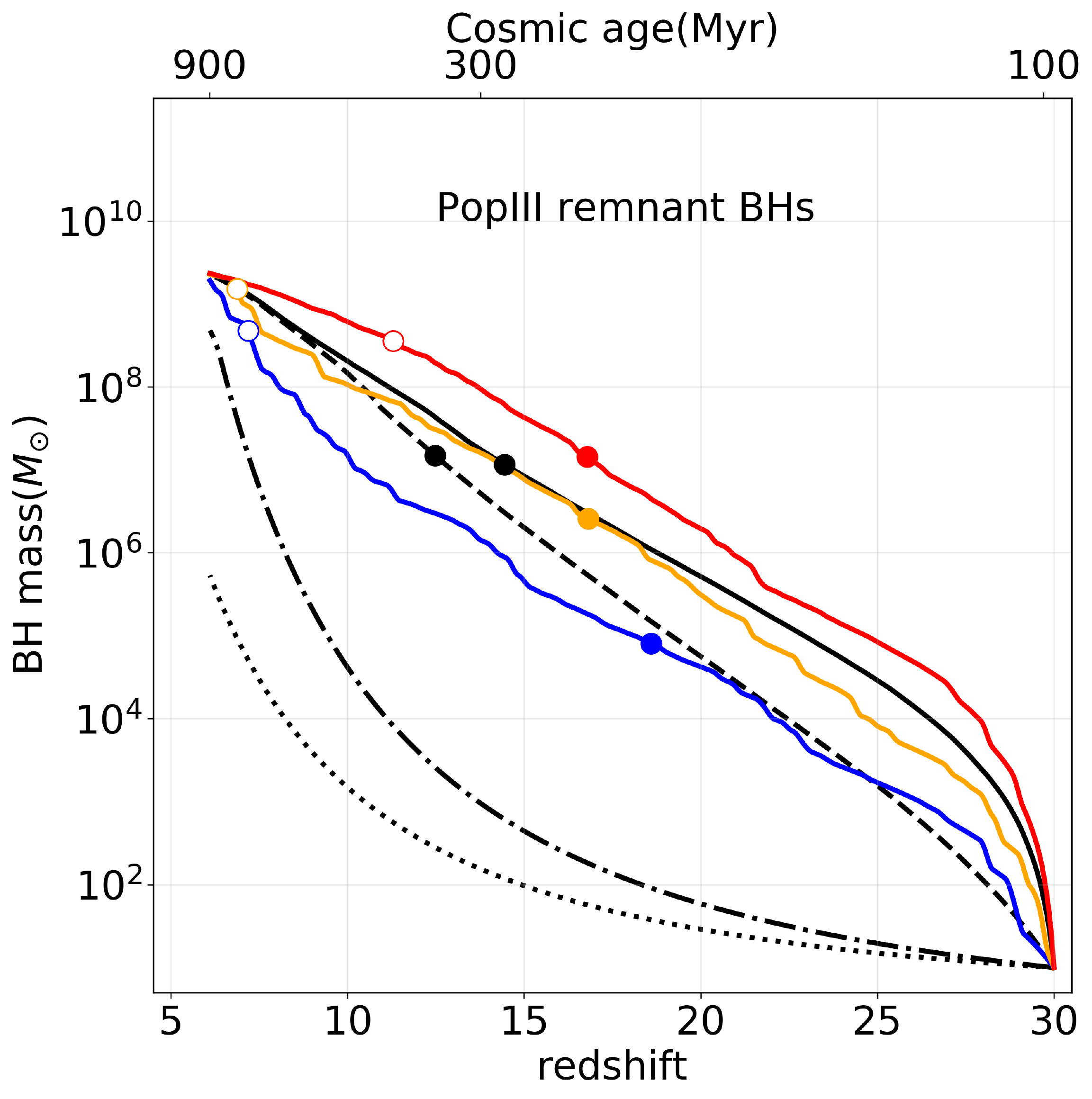}\hspace{10mm}
\includegraphics[scale=0.34]{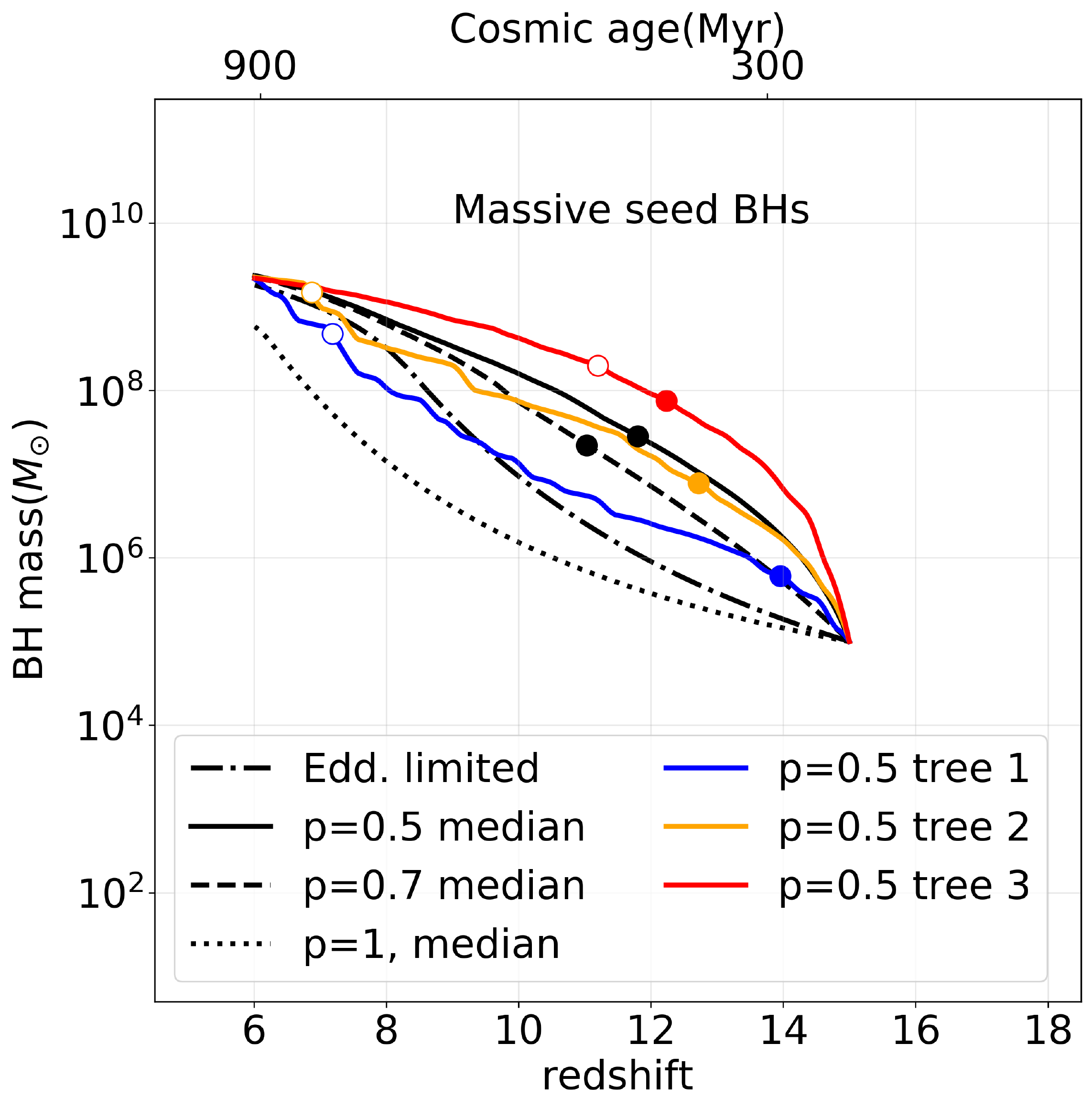}
\caption{Evolutionary tracks of accreting BHs in two seeding scenarios: PopIII remnant BHs with $M_{\rm BH,0}=10~M_\odot$ 
at $z=30$ (left panel) and massive seed BHs with $M_{\rm BH,0}=10^5~M_\odot$ at $z=15$ (right panel). The galactic inflow 
rates are taken from Fig~.\ref{fig:mergertree} ($\mathcal{F}=0.1$). In the fiducial case ($p=0.5$; solid), the BH mass 
in all the cases (three representative trees and median tree) converge to $M_{\rm BH}\sim10^9~M_\odot$ by $z=6$, while 
the evolutionary tracks show a great diversity in the earlier stage depending on the halo merger assembly process. With 
higher $p$ values $(\geq\, 0.7)$, stronger outflows delay or even suppress BH growth. As a reference, the Eddington-limited 
growth curve is overlaid with the black dashed-dotted curve. 
The filled and open circle on each curve marks when the BH feeding rate first falls below and last exceeds the 
Eddington accretion rate, respectively. 
In the epoch between the two circles, the BH grows via multiple intermittent super-Eddington accretion modes.
Note that there is a clear transition between the two phases on the median tree.
}
\vspace{5mm}
\label{fig:growth_model}
\end{figure*}

The mass inflow rate from galactic scales $\dot{M}_0$ is a parameter determined by the environment where the BH is hosted.
Here, we estimate this value as the baryonic mass growth rate of a massive dark matter (DM) halo that ends up as a high-$z$ quasar host galaxy
with mass of $M_{\rm h}\sim 10^{12}~M_{\odot}$ at $z\gtrsim 6$.
Following \citet{Li2021}, we construct merger trees to track the growth of the DM halos in highly-biased, overdense regions of the universe
and plant a seed BH with $M_{\rm BH,0}$ at $z=z_0$ in each tree.

In Fig.~\ref{fig:mergertree}, we present the baryonic mass inflow rate into the progenitor DM halos defined by 
$\dot{M}_{\rm h}(\Omega_{\rm b}/\Omega_{\rm m})$ for $10^4$ different trees obtained in \citet{Li2021}, where 
$\Omega_{\rm b}/\Omega_{\rm m}=0.156$ is the baryon fraction \citep{Planck2018}. 
We highlight three representative trees (red, orange, and blue) and overlay the median value of the mass growth rate (black).
Note that the inflow rate shown in Fig.~\ref{fig:mergertree} is considered to be an upper bound for the gas supply rate to the nuclei
because a certain fraction of the gas is reduced owing to various effects, e.g., angular momentum transport of inflowing gas
and gas consumption by star formation.
Taking into account these effects, we specify the mass inflow rate to the galactic center as
\begin{equation}
\dot{M}_0(z)=\mathcal{F}\cdot \frac{\Omega_{\rm b}}{\Omega_{\rm m}}~\dot{M}_{\rm h}(z),
\end{equation}
and treat the value of $\mathcal{F}$ as a free parameter.
It is worth noting that the efficiency factor is at most $\mathcal{F}\la 0.1$ without star formation prescriptions \citep{Hopkins2010}.
Thus, we demonstrate the impact of this choice on BH growth, with the restriction of $\mathcal{F}\leq 0.1$.

In the following discussion, we focus on two BH seeding models:
(i) $M_{\rm BH,0}=10~M_\odot$ at $z_0=30$ and
(ii) $M_{\rm BH,0}=10^5~M_\odot$ at $z_0=15$. 
The former case corresponds to a remnant BH originating from a first-generation star (Population III star, hereafter PopIII star), 
while the latter case mimics a heavy BH seed through supermassive star formation in massive DM halos under 
peculiar environments \citep{Dijkstra2008, Inayoshi2018, wise2019formation,Lupi2021}.
A semi-analytical study by \citet{Li2021} suggests that the formation of BH seeds is rather promoted in the high-$z$ quasar 
progenitor halos located in the overdense regions and yields a BH mass distribution ranging from several hundred to above $10^5~M_{\odot}$.
We study the two scenarios that bracket the low and high mass ends for BH seeds.
However, note that dense, metal-poor environments also allow the formation of BH seeds in the the intermediate mass range 
through stellar collisions \citep{Sassano2021,Tagawa2021}.

\section{The Growth of Seed BHs}\label{sec:seed}

\begin{figure*}
\centering
\includegraphics[scale=0.34]{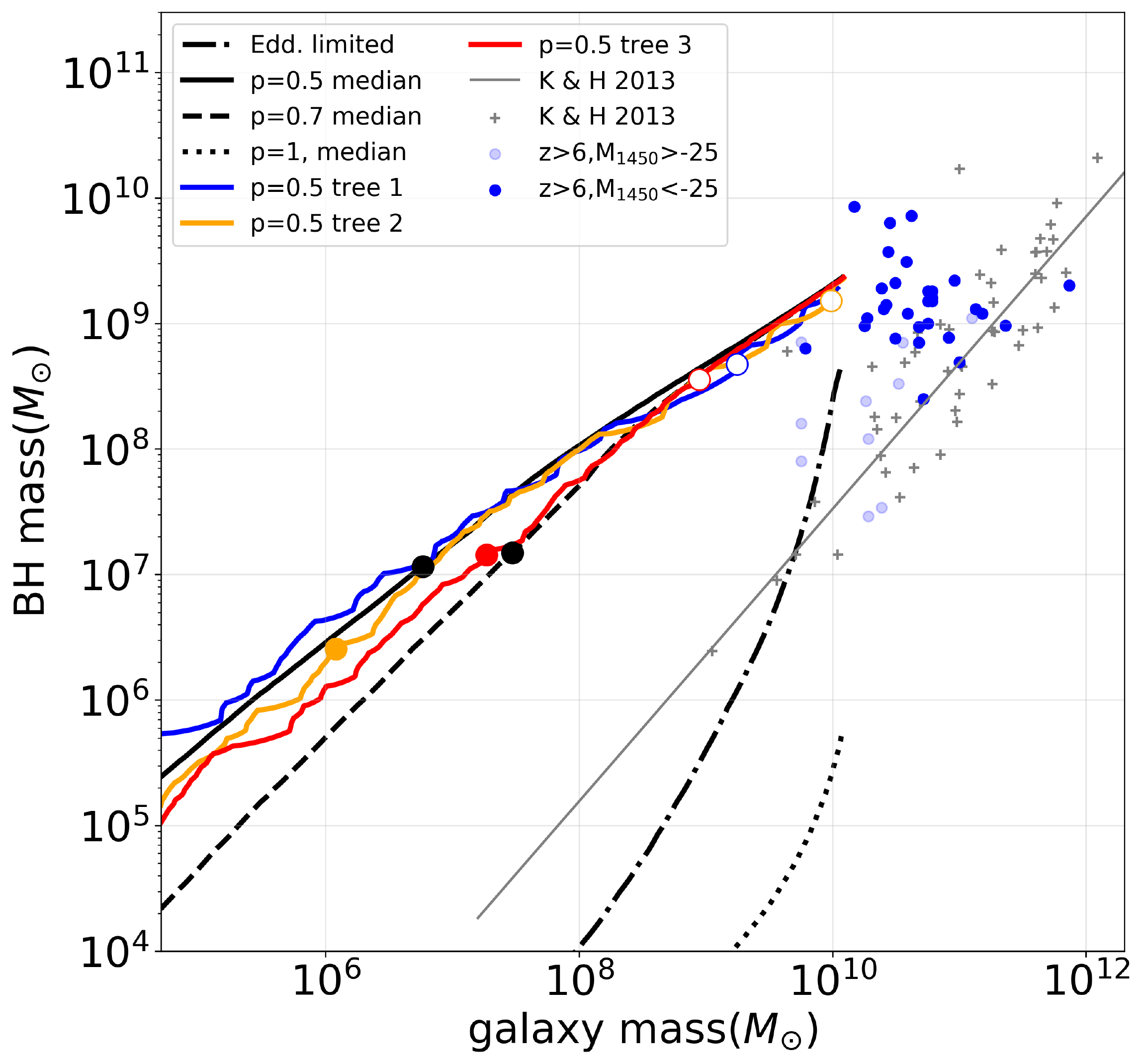}\hspace{10mm}
\includegraphics[scale=0.34]{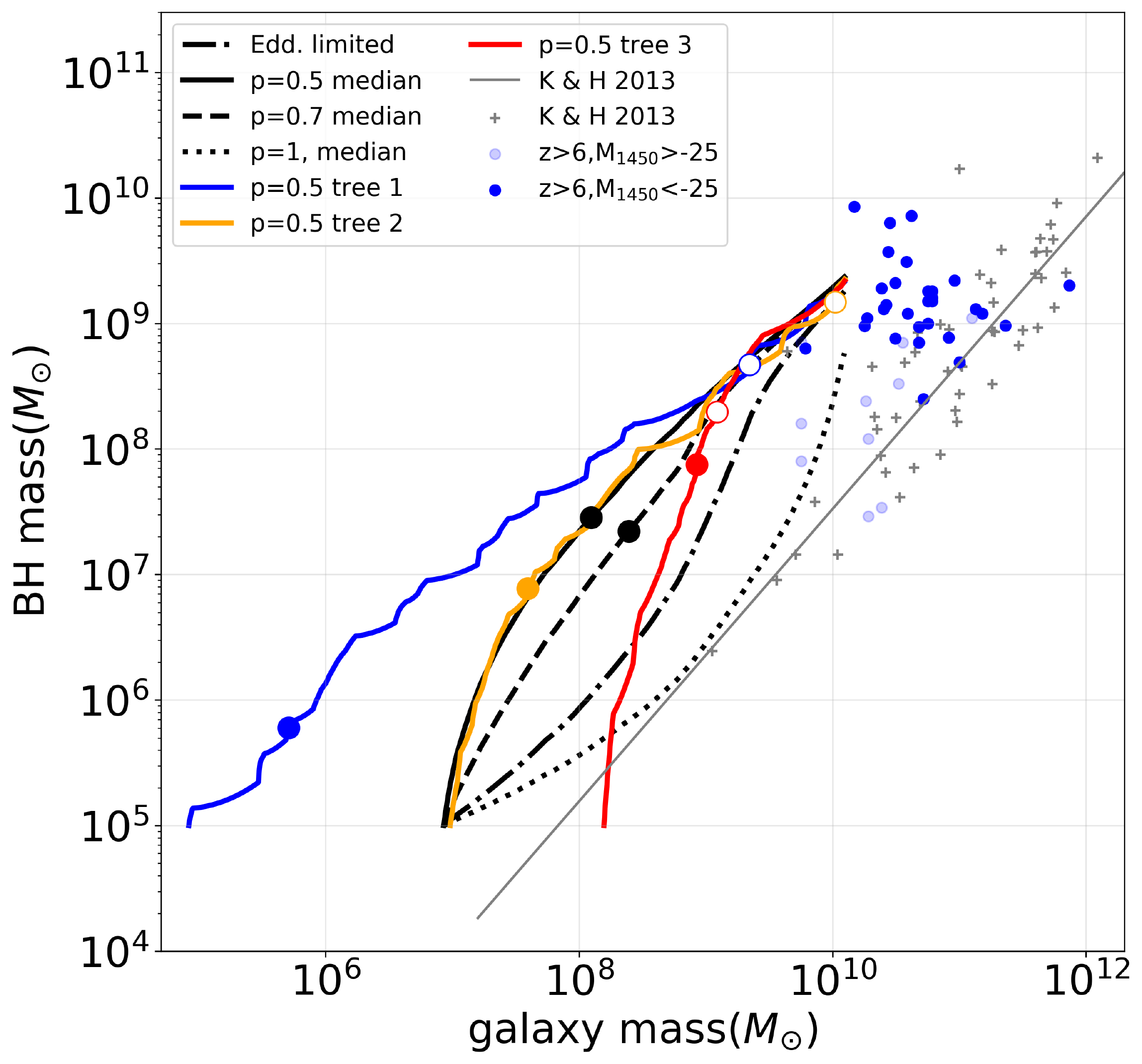}
\caption{Coevolution diagram for growing seed BHs and their host galaxies for the two scenarios; PopIII remnant BHs (left panel, $6\la z\la 30$) 
and massive seed BHs (right panel, $6\la z\la 15$). The curves correspond to those shown in Fig.~\ref{fig:growth_model}. 
The galaxy mass is calculated with the stellar/halo mass ratio in \citet[][]{Behroozi2019}. 
The blue symbols show the high-$z$ quasar samples compiled by \citet{Izumi2019}.
Note that the galaxy mass is calculated from the [\ion{C}{2}]-based dynamical mass using a conversion factor calibrated in low-$z$ galaxies \citep{Tacconi2018}. 
For reference, the local observational data and best-fit relation are overlaid \citep[grey crosses and solid line;][]{kormendy2013coevolution}. 
The BH growth model yields a high BH-to-galaxy mass ratio above the local relation even in the presence of strong outflows.
}
\vspace{5mm}
\label{fig:MM_plot}
\end{figure*}

Fig.~\ref{fig:growth_model} shows the growth history of a BH seed with an initial mass
of $M_{\rm BH,0}=10~M_\odot$ at $z=30$ (left) and $M_{\rm BH,0}=10^5~M_\odot$ at $z=15$ (right), respectively.
For the fiducial case with $p=0.5$, we show four cases, with the galactic mass inflow rate given by the curves
highlighted in Fig.~\ref{fig:mergertree}.
To demonstrate the impact of outflow strength, we show two alternative cases with
$p=0.7$ (dashed) and $p=1.0$ (dotted) along with the median tree.
For comparison, the Eddington-limited growth curve with a 100\% duty cycle is overlaid (dashed-dotted).

In the fiducial cases, both light and heavy BH seeds grow at super-Eddington rates in the earlier epoch 
since the galactic inflow rate $\dot{M}_0$ exceeds the Eddington value substantially and thus
the net accretion rate is kept as high as $\ga 10~\dot{M}_{\rm Edd}$ even with strong outflows.
The history shows a great diversity depending on the halo merger assembly process, but the BH mass converges 
to $M_{\rm BH}\simeq 2\times 10^9~M_\odot$ by $z\simeq 6$. Continuous super-Eddington accretion is sustained 
down to $z\sim17$ and $z\sim12$ (filled circles) for the light and heavy seed scenario, respectively, and the accretion behavior 
turns into multiple intermittent phases at lower redshifts. The overall trend of BH growth is consistent with that found 
in a previous semi-analytical model by \citet{Pezzulli2016}, where the transition redshift is as low as $z \sim 10$.
For the case with $p=0.7$, stronger outflows reduce the net accretion rate more significantly in the early stage.
However, when the BH mass is high enough that the galactic inflow rate is below the Eddington rate, 
the outflow effect plays a less important role in suppressing the BH growth.
As a result, all the cases with $p\simeq 0.5-0.7$ yield a comparable BH mass at $z=6$.

In contrast, for the case with the strongest outflow ($p=1.0$), mass growth of PopIII remnants is quenched at $z>15$ and 
the mass reaches only $\sim 10^6~M_\odot$ by $z\simeq 6$, while the heavy seed BH reaches 
$\sim 5\times 10^8~M_\odot$.
The result clearly shows less-massive seed BHs tend to be significantly affected by mass loss via outflows
when the suppression effect is significant ($p\ga 0.7$). For comparison, a previous study by \citet{Madau2014} discusses 
that PopIII remnant BHs grow to be SMBHs via mildly super-Eddington accretion ($\sim 3~\dot{M}_{\rm Edd}$) in the absence 
of mass loss through strong outflows. The presence of outflows changes the growth history of seed BHs at higher redshifts.
The outflow strength $p$ is constrained by various simulations suggesting $p\sim0.5-1$ \citep[][references therein]{Yuan2014}. 
However, a relatively small value of $p\simeq 0.5$ is required to explain the existence of high-$z$ SMBHs in our models. 
As discussed in \S\ref{sec:method}, the outflow strength should depend primarily on the response of the BH to its surrounding environments.
It is of significant importance for future observations to constrain the outflow strength in order to better understand interactions
between central SMBHs and host galaxies.

\begin{figure*}
\centering
\includegraphics[scale=0.3]{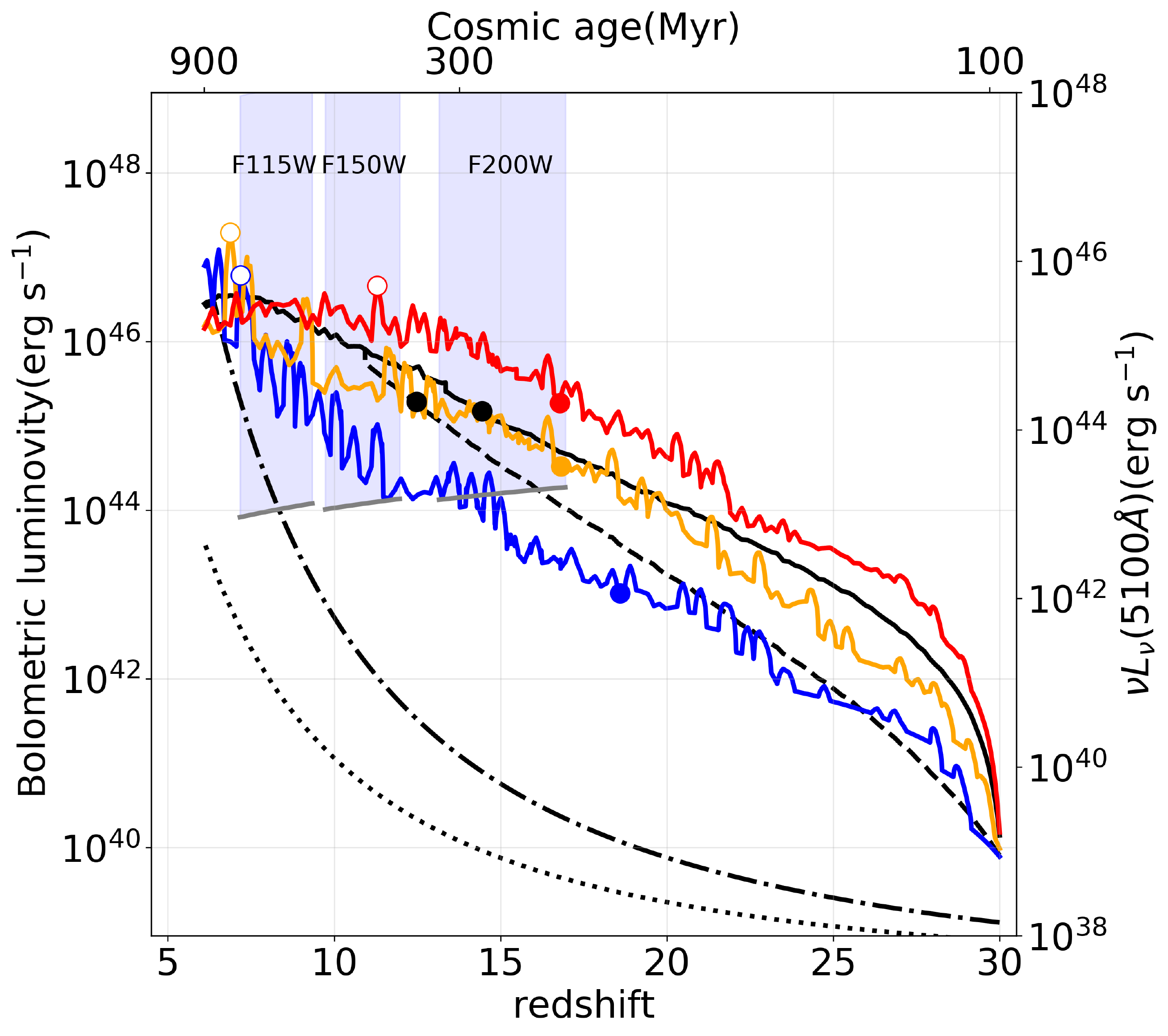}\hspace{10mm}
\includegraphics[scale=0.3]{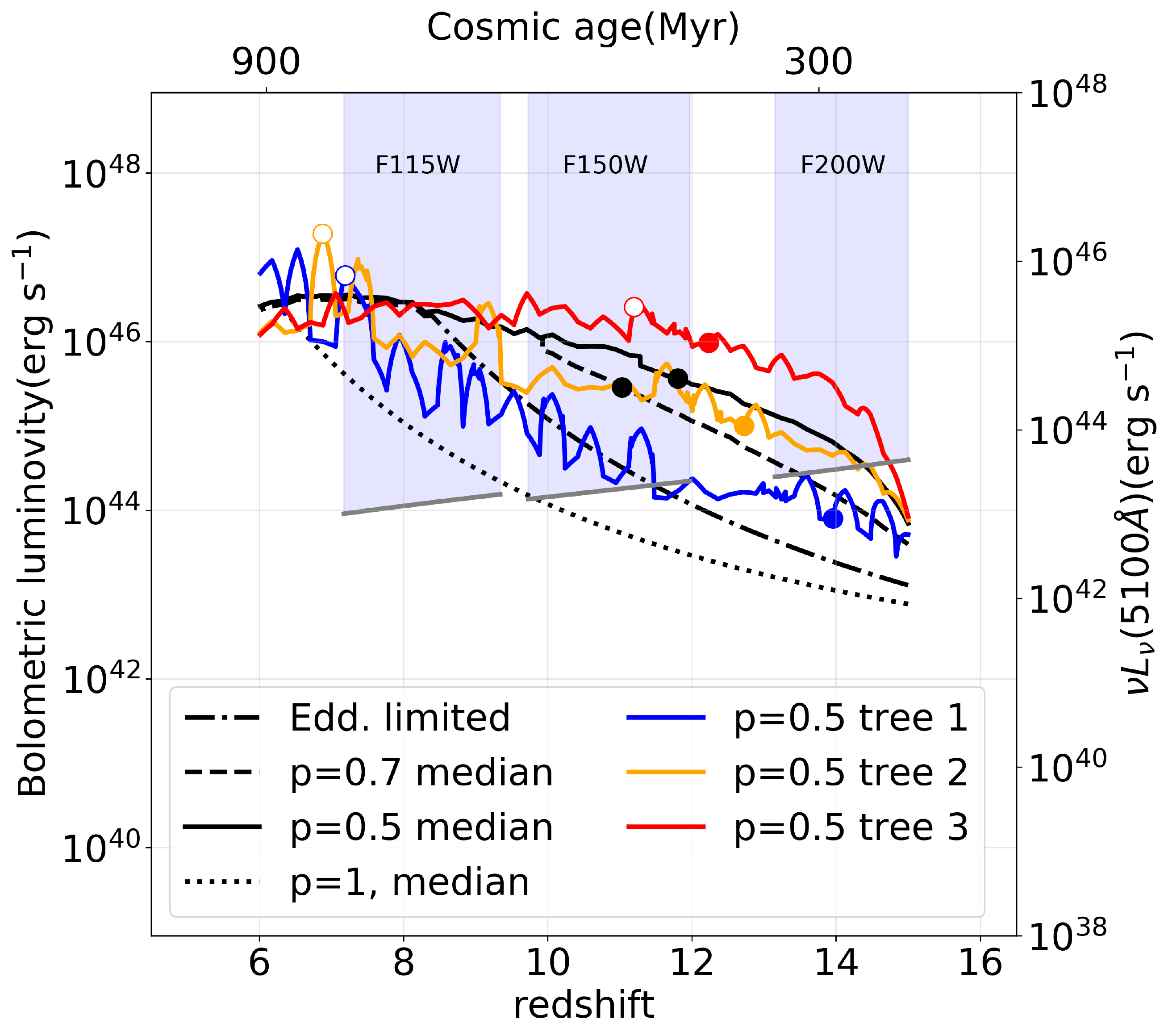}
\caption{Evolution of the bolometric luminosity evolution for PopIII remnant BHs (left panel) and massive seed BHs (right panel). 
The curves correspond to those shown in Fig.~\ref{fig:growth_model} and the luminosity is calculated with Eq.~(\ref{eq:bol_lum}).
The luminosity at $\lambda = 5100~{\rm \AA}$ calculated with $\nu L_{\nu}(5100~{\rm \AA})=L_{\rm BH}/9.0$ \citep{Kaspi2000}
is shown as reference for single-epoch BH mass measurements with H$\beta$ line emission.
The flux limits of JWST/NIRCam with the F115W, F150W, and F200W filter are overlaid with the the grey curves (see Eq.~\ref{eq:limit_lum}).
}
\vspace{5mm}
\label{fig:Lum_z}
\end{figure*}

Fig.~\ref{fig:MM_plot} presents the evolutionary track of the BH-to-galaxy mass ratio.
Here, we calculate the mass of the host galaxy as $f_\star M_{\rm h}$, by assuming a stellar-to-halo mass ratio
$f_\star(z=6,M_{\rm h})\sim 0.002-0.015 $ \citep[see Eqs.~J1 -- J8 in][]{Behroozi2019}.
We note that the choice of $f_\star = 0.01$ corresponds to a conversion efficiency from gas into stars, 
$\epsilon_\star \simeq 0.05$, which is motivated by abundance matching and the observed UV luminosity function 
of galaxies at $z\simeq 6$ \citep{Bouwens2015}.
Under this assumption, the BH mass grows faster than the host galaxy mass does, leading to a BH-to-galaxy mass ratio of 
$M_{\rm BH}/M_\star \simeq 0.1$.
The mass ratio approaches a constant value of $0.1$ by $z=6$, which is $\ga 10\sim 100$ times higher than the local empirical relation
\citep{kormendy2013coevolution} but is consistent with those of $z >6 $ \citep{Pensabene2020}. 
Additionally, the existence of such overmassive BHs in protogalaxies will provide us with a unique opportunity of detecting highly 
accreting seed BHs in the very early universe at $z > 10$ by upcoming observations, e.g., the 
{\it James Webb Space Telescope} (JWST) and {\it Nancy Grace Roman Space Telescope} (RST) \citep{Inayoshi2022}.

Finally, we briefly mention the dependence on the choice of $\mathcal{F}$, which characterizes the reduction of the mass inflow rate
from galactic scales to the nuclear regions \citep[e.g.,][]{Hopkins2010}. 
Overall, for $10^{-3}\leq \mathcal{F} \leq 0.1$, the BH mass at $z\simeq 6$ is proportional to $\mathcal{F}$, regardless of the seeding models.
 Namely, the final mass is approximated as $M_{\rm BH} \simeq 2\times 10^9 (\mathcal{F}/0.1)~M_{\odot}$ based on the median tree with $p=0.5$.
This is because for the majority of the cosmic time, the seed BHs  grow at 
super-critical rates---nevertheless, the majority of the final BH mass is accreted mostly via sub-critical growth at 
$\dot{M}_{\rm BH}=\dot{M}_0\propto \mathcal{F}$, during the last few e-foldings (as indicated by dots in curves in Fig.~\ref{fig:growth_model}).
We note that this scaling is also applied to the BH-galaxy coevolution shown in Fig.~\ref{fig:MM_plot}.

Observationally, the efficiency of gas feeding from galactic scales down to the nuclear regions of high-$z$ quasars
has been poorly constrained.
In fact, the highest spatial resolution of submillimeter observations with the Atacama Large Millimeter Array (ALMA)
enables us to address the central regions at $\la 1~{\rm kpc}$ for $z\sim 6$ quasar 
host galaxies\footnote{https://public.nrao.edu/telescopes/alma/} \citep{Venemans2019,Walter2022}. 
Given the existence of overmassive SMBHs at the high-$z$ universe, it is plausible that a large fraction of gas feeds the nuclear 
SMBH at cosmic dawn, though detailed physical processes to maintain $\mathcal{F} \simeq 0.1$ have been poorly understood.
We leave the constraints on $\mathcal{F}$ to future observations and simulation studies.

\section{Discussion} \label{sec:discussion}

Upcoming observations with JWST will provide a unique opportunity to detect fast-accreting seed BHs 
that offer an evolutionary pathway toward the overmassive population over their host galaxies at $z>6$.
Fig.~\ref{fig:Lum_z} shows the bolometric luminosity produced by BHs for the two seeding scenarios;
light seeds (left) and heavy seeds (right). 
Along the halo assembly history, the bolometric luminosity is calculated as
\begin{equation}
\frac{L_{\rm  BH}}{L_{\rm Edd}}=
\left\{
\begin{array}{ll}\dot{m} & (\dot{m}<2), \vspace{3mm}\\ 
2\left[1+\ln \left(\frac{\dot{m}}{2}\right)\right] & (\dot{m} \geq 2),
\end{array}
\right.
\label{eq:bol_lum}
\end{equation}
 \citep{Watarai2001}, where $\dot{m}\equiv \dot{M}_{\rm BH}/\dot{M}_{\rm Edd}$ is the dimensionless BH feeding rate. 
We note that this formula is consistent with the results from various RHD simulations (\citealt{Ohsuga2005,jiang2014,sadowski2015};
see also Fig.~5 in \citealt{Inayoshi2020}).
For both seeding scenarios, the bolometric luminosity is as low as $\la10^{44}~{\rm erg}$ s$^{-1}$ at earlier epochs even though the BHs grow at 
supercritical rates. When the BH mass exceeds several $10^6~M_{\odot}$, nearly independent of the seed model and the host halo assembly 
history, the bolometric luminosities become as high as $L_{\rm bol} \simeq 10^{45-46}~{\rm erg~s}^{-1}$, 
which is reachable with wide-field deep surveys such as the Hyper Suprime-Cam (HSC) Subaru Strategic Program \citep{Matsuoka2018b,Onoue2019}.

The detection limit is improved by more than one order of magnitude with the deeper JWST observations.
Here, assuming the radiation spectra of seed BHs to be a broken power-law consistent with the stacked UV spectra 
of quasars at $z\simeq 2.4$ \citep{Lusso2015}, the luminosity density at a characteristic frequency of $\nu_0=10~{\rm eV}/h$ is given by
$L_{\nu_0}=L_{\rm BH}/(\alpha \nu_0)$, where $\alpha=3.3$ \citep[see also][]{Inayoshi2022}. 
Therefore, for a given detection threshold of a JWST NIRCam filter, the critical bolometric luminosity is calculated by
\begin{equation}
L_{\rm crit} = 4\pi D_{\rm L}^2~ \alpha \nu_{\rm obs} F^{\rm jwst}_{\nu_{\rm obs}},
\label{eq:limit_lum}
\end{equation}
where $\nu_{\rm obs}=\nu_0/(1+z)$, $D_{\rm L}$ is the luminosity distance, and $F^{\rm jwst}_{\nu_{\rm obs}}$ is 
the limiting flux of a NIRCam filter that covers the corresponding frequency.
For reference, we overlay the signal-to-noise ratio (S/N) = 10 detection limit of JWST/NIRCam imaging in a 10k second exposure
with the F115W, F150W, and F200W filter\footnote{https://jwst-docs.stsci.edu/jwst-near-infrared-camera/nircam- instrumentation/nircam-filters}, 
which cover a redshift range of $7\la z\la17$.
For the PopIII remnant scenario, the accreting BHs can be detectable up to $z\sim17$ for the fast growing case (red curve) 
even with a lower value of $\mathcal{F}\sim 0.01$, leading to nearly 10-fold reduction of the bolometric luminosity. 
On the other hand, massive seed BHs require a high value of $\mathcal{F}\simeq 0.1$ (our fiducial choice) to be detectable at 
$z\gtrsim 13$ with the F200W filter, while the highest observable redshift is limited to $z\lesssim 10-12$ with a lower value of $\mathcal{F}\sim 0.01$.
In conclusion, it is plausible but dependent sensitively on the value of $\mathcal{F}$ 
that seed BHs can be observable even at $z\gtrsim 13$, where the rapid growth of the DM halo facilitates BH growth even in the presence of outflows.
The expected detection number of such accreting seed BHs is one in ten NIRCam fields of view
at the depth of 10 ks exposures \citep{Inayoshi2022}
with the help of photometric selection for seed BH candidates \citep[e.g.,][]{Pacucci2016,Natarajan2017,Valiante2018,Inayoshi2022b}
as well as the measurement of the BH-to-galaxy mass ratio \citep{Scoggins2022}.
\citet{Inayoshi2022b} recently found that seed BHs growing at super-Eddington rates produce extremely strong Balmer lines 
(the rest-frame H$\alpha$ equivalent width is $\simeq 7$ times larger than the typical vale for low-$z$ quasars) because of efficient collisional excitation of hydrogen
to higher levels ($n\geq 3$) in the dense disk.
The broadband color is redder owing to strong H$\alpha$ emission and thus the multiband photometry with NIRCam and MIRI 
enables us to robustly select this extremely young BH at $z\sim 7-12$.
For reference, we also show the luminosity at $\lambda=5100$ \AA, which is used for single-epoch BH mass measurements with H$\beta$ line emission.

We note that the accretion model in Eq.~(\ref{eq:inflow}) characterizes suppression of BH feeding owing to mass loss via outflows, but does not 
take into account the impact of mechanical feedback on large galactic scales. With the prescription given by Eq.~(20)-(24) in \citet{Hu2022}, the total kinetic energy produced 
by nuclear BHs during continuous supercritical accretion (from the seeding redshift to the transition epoch denoted with filled circles) is 
as high as $E_{\rm kin} \sim 10^{58}~{\rm erg}~(3\times 10^{57}~{\rm erg})$ for the PopIII seed (massive seed) scenario. Along the median halo tree, the binding energy 
for gas within the virial radius at the transition epoch is $E_{\rm b,gas} \sim 8\times 10^{56}~{\rm erg} ~(9\times 10^{55}~{\rm erg})$. Therefore, mechanical feedback associated 
with fast-growing BHs would affect the structure of galactic inflows if the efficiency of energy deposition into the gas is sufficiently high; 
$\epsilon \gtrsim 7.9\% ~(2.9\%)$. In fact, the feedback efficiency depends sensitively on the energy loss via radiation  \citep[$\epsilon \sim 1\%$;][]{Kitayama2005} 
and the geometry, dynamical and thermodynamic state of the gaseous medium through which the outflow propagates 
\citep{Costa2014}. To explore these effects is left for future investigation.

The BH growth model presented here focuses on highly biased regions of the 
universe where high-$z$ quasar hosts harbored in massive DM halos with $M_{\rm h}\sim 10^{12}~M_{\odot}$ form by $z\sim 6$.
Such massive halos are expected to be ideal sites for the formation of massive seed BHs owing to their peculiar environments,
e.g., strong ultra-violet irradiation from nearby galaxies and violent galaxy mergers \citep[also see][]{Li2021,Lupi2021}
and sufficiently feed the central regions via intense cold gas streams \citep{DiMatteo2012}.
We find that the galactic gas inflows triggered during their assembly promote rapid growth of seed BHs in the protogalactic nuclei,
nearly independent of their initial mass. 
This mechanism naturally explains the puzzling appearance of $>10^9~M_{\odot}$ BHs at $z>6$. 
However, the quick assembly of seed BHs via super-Eddington mass accretion would make it difficult to distinguish 
their seeding models through the BH-to-galaxy mass ratio \citep{Visbal2018}.

\section{Summary} \label{sec:summary}

In this paper, we propose an assembly model for the SMBHs that end up in rare massive $\sim 10^{12}~M_\odot$ host halos at $z \sim 6-7$, 
applying a kinetic feedback prescription for BHs accreting above the Eddington rate, provided by RHD simulations for the long-term evolution 
of the accretion-flow structure \citep{Hu2022}.
We incorporate this feedback model for two different BH seeding scenarios: PopIII stellar remnant BHs and massive seed BHs.
For each case, we study the evolutionary pathway of the BH-to-galaxy mass ratio and the detectability of fast growing BHs
with upcoming JWST observations.

The large inflow rates into those high-$z$ quasar progenitor halos during their assembly enable the formation of $>10^9~M_\odot$ SMBHs by $z\sim 6$, 
even starting from PopIII remnant BHs at $z\sim 30$, and even in the presence of outflows that reduce the BH feeding rate.
This overall trend holds for both seeding models when mass loss associated with outflows from super-Eddington accretion flows
is moderate; namely $p\simeq 0.5$, where the value of $p$ characterized the reduction of the mass inflow down to the BH (see Eq.~\ref{eq:inflow}). 
Stronger outflows with $p\ga 0.7$ reduce the BH mass achievable from PopIII remnant BHs substantially but give a smaller impact on
the growth of massive seed BHs,
For the cases where seed BHs grow to be SMBHs by $z\sim 6$, those BHs tend to be overmassive relative to their host galaxies 
compared to the local relations and show a high BH-to-galaxy mass ratio of $\sim 0.01$, despite different strength of outflows. 

The high luminosity from those growing BHs makes themselves detectable with upcoming JWST observations. 
In the most optimistic (fastest growing) case, JWST can reveal the rapid growing BHs up to redshift $z\sim17$ (red curve in Fig.~\ref{fig:Lum_z}). 
However, multiple diagnostics such as color-color selection, spectral analysis as well as measurements of the BH-to-galaxy mass ratio
are required to identify the nature and origin of the fast growing seeds. 
Meanwhile, the rapid assembly phases of BHs also make it difficult to distinguish whether those BHs originate from
PopIII stellar remnants or massive seed BHs.

As a caveat, the BH assembly history in our model depends sensitively on the outflow strength $p$ and efficiency of gas feeding from 
galactic to nuclear regions $\mathcal{F}$, both of which are poorly constrained by observations. 
Therefore, it is crucial for future observations and theoretical studies to shed light into 
these assumptions for our better understanding of the interaction between nuclear BHs and host galaxies.

\begin{acknowledgements}
We thank the anonymous referee for a careful reading of our manuscript and comments that helped 
improve this paper. We thank Takuma Izumi, Masafusa Onoue, and Ran Wang for useful discussions.
K.I. acknowledges support from the National Natural Science Foundation of China (12073003, 
11721303, 11991052, 11950410493), the National Key R\&D Program of China (2016YFA0400702), 
and the China Manned Space Project with NO. CMS-CSST-2021-A06. Z.H. acknowledges support 
from NSF grant AST-2006176. 
R.K. acknowledges financial support via the Heisenberg Research Grant funded by the German Research 
Foundation (DFG) under grant NO. $\sim$KU 2849/9. Further, R.K. acknowledges financial support via the JSPS 
Invitational Fellowship for Research in Japan under the Fellowship ID S20156.
The numerical simulations were performed with the Cray XC50 
at the Center for Computational Astrophysics (CfCA) of the National Astronomical Observatory of 
Japan and with the High-performance Computing Platform of Peking University.
\end{acknowledgements}



\end{CJK*}
\end{document}